\documentclass[aps,prc,floatfix,twocolumn,showpacs,tightenlines,
preprintnumbers,superscriptaddress]{revtex4}
\usepackage{amsmath,amssymb,amsfonts,mathrsfs}
\usepackage{dcolumn}
\usepackage{supertabular}
\usepackage{color,graphicx}
\usepackage[bookmarksopen]{hyperref}
\newcommand{\bm}{\bibitem}

\allowdisplaybreaks

\def  \bc   {\begin{center}}
\def  \ec   {\end{center}}
\def  \ba   {\begin{array}}
\def  \ea   {\end{array}}
\def  \be   {\begin{equation}}
\def  \ee   {\end{equation}}
\def  \bea  {\begin{eqnarray}}
\def  \eea  {\end{eqnarray}}

\def \bla#1\ela {\begin{flalign}#1\end{flalign}} 

\def  \bd   {\begin{displaymath}}
\def  \ed   {\end{displaymath}}
\def  \bse  {\begin{subequations}} 
\def  \ese  {\end{subequations}}
\def  \bwt  {\begin{widetext}}
\def  \ewt  {\end{widetext}}
\def  \nn   {\nonumber}

\def  \pio  {\pi^0}
\def  \mn  {\widetilde{M}}
\def  \gn  {\widetilde{G}}
\def  \en  {\widetilde{E}}
\def  \vk   {{\bf k}}
\def  \vq   {{\bf q}}
\def  \vr   {{\bf r}}
\def  \g    {\mathrm{ g}}
\def \gpim {\frac{\mathrm{g}_\pi}{2M_N}}
\def \getam {\frac{\mathrm{g}_\eta}{2M_N}}

\def \gam {\frac{\mathrm{g}_a}{2M_N}}
\def \gbm {\frac{\mathrm{g}_b}{2M_N}}

\renewcommand{\sl}[1]{{#1}\!\!\!/}

\begin{document}
\title{Effect of in-medium $\pi$ and $\eta$ propagators to charge symmetry breaking interaction}
\author{Subhrajyoti Biswas}
\email{anjansubhra@gmail.com}
\affiliation{Department of Physics, Rishi Bankim Chandra College, Naihati - 743165, West Bengal, India}
\medskip
\begin{abstract}
We revisit the charge symmetry breaking $(CSB)$ in nucleon-nucleon $(NN)$ interaction 
caused by $\pi$-$\eta$ mixing in nuclear matter $(NM)$ employing one boson exchange
$(OBE)$ model. The medium effect on $CSB$ is incorporated through the in-medium 
$\pi$ and $\eta$ propagators. In addition, the medium modification to the nucleon mass 
is also taken into account to construct the $CSB$ class $III$ potential considering 
off-shell $\pi$-$\eta$ mixing amplitude and large contribution of in-medium meson propagators 
to $CSB$ potential is found for the pseudoscalar $(PS)$ interaction compared to that of pseudovector 
$(PV)$ interaction.
\end{abstract}
\vspace{0.08 cm}
\maketitle
\section{Introduction}
The charge symmetry $(CS)$ is broken inherently by small amount in the nucleon-nucleon 
$(NN)$ interaction \cite{Henley69, Henley79, Machleidt89, Miller90,  Miller95}. This symmetry 
breaking effect is observed trivially in the neutron-neutron $(nn)$ and proton-proton 
$(pp)$ systems through the presence of the Coulomb interaction. However, it is difficult 
to separate the strong interaction part model independently from the Coulomb interaction   
\cite{Henley69, Henley79, Machleidt89, Miller90, Miller95}.  

At the fundamental level $CS$ of the nuclear force is broken due to the down $(d)$ and up $(u)$ 
quark mass difference {\em i.e.} $m_d \neq m_u $ and the electromagnetic interactions among 
the quarks \cite{Henley69,Henley79, Machleidt89, Miller90, Miller95}. The $d$ and $u$ quarks 
mass difference along with the electromagnetic effects is responsible for the observed mass 
differences between hadrons of the same isospin multiplets. Such mass splitting causes $CSB$ 
at the hadronic level \cite{Henley69,Henley79, Machleidt89, Miller90, Miller95}.  

Many experiments have been designed to detect and measure $CSB$ effects in 
various observables \cite{Miller86, Williams87, Ge87, Gersten88, Holzenkamp87, 
Niskanen88, Iqbal87, Iqbal88}. It is seen clearly in the small difference between the 
$^1S_0$ $nn$ and $pp$ scattering lengths. The latest data of scattering 
experiment show that the amount of $CSB$ is $\Delta{a}_{CSB} = a^N_{pp}- a^N_{nn} = 
1.6\pm0.6~\mathrm{fm}$ \cite{Miller90, Howell98, Gonzalez99,Coon87} where the 
superscript $N$ indicates the nuclear effect only.
 
Another convincing observation of $CSB$ is found in the difference of ground state 
binding energies between the mirror nuclei $^3He$ and $^3H$. After excluding the 
corrections due to the static Coulomb interaction $(648\pm4~\mathrm{KeV})$ 
\cite{Brandenburg88, Friar87, Wu90}, electromagnetic effect $(35\pm3~\mathrm{KeV})$ 
\cite{Wu90,Brandenburg78} and $n$-$p$ mass difference in the kinetic energy 
$(14\pm2~\mathrm{KeV})$ \cite{Friar90}, the remaining $67\pm9~\mathrm{KeV}$
is believed to be accounted for the $CSB$ interaction. 

Similar phenomena have been investigated for other mirror nuclei \cite{Okamoto64,Nolen69,Epele92}. 
The Coulomb displacement energies of mirror nuclei are found different. This is known 
as the Okamoto-Nolen-Schiffer $(ONS)$ anomaly \cite{Okamoto64,Nolen69,Epele92}. Many efforts, 
considering electromagnetic corrections, many-body correlations etc. have been made to explain 
this anomaly \cite{Auerbach72, Sato76, Blunden87, Tam81}. In addition other manifestation of 
$CSB$ interaction are the difference of $n$-$p$ form factors, correction to $\mathrm{g}-2$ etc. 
\cite{Miller06}.  

The well known mechanism that generates $CSB$ nucleon-nucleon $(NN)$ interaction 
is the mixing of neutral mesons with different isospins but same spin and parity 
like $\pi$-$\eta$, $\pi$-$\eta^\prime$, $\rho$-$\omega$ etc. mixing. Such mixing
of isospin pure resonance states is caused by the $d$-$u$ quark mass difference 
and electromagnetic interactions \cite{Henley69, Henley79, Miller95}. In a quark 
model calculation \cite{Goldman92}, Goldman, Henderson, and Thomas showed that 
$\rho$-$\omega$ mixing amplitude has a substantial momentum dependence employing 
free constituent quark propagators and phenomenological meson-quark-antiquark vertex 
form factors. A $QCD$ sum rule calculation \cite{Hutsuda94} and other investigations 
\cite{Krein93, Williams91, Burden92, Roberts92, Mitchell93, Mitchell94, Piekarewicz93, 
Oconnell94, Oconnell97} also reported strong momentum dependence of $\rho$-$\omega$ mixing. 

Using chiral perturbation theory Maltman shows significant change of $\pi$-$\eta$ mixing 
amplitude in going from timelike to spacelike $q^2$ \cite{Maltman93}. Similar $q^2$
dependence is found at the leading order contribution of $\pi$-$\eta$ mixing obtained 
from chiral effective Lagrangian model \cite{Manohar84,Gasser85A, Gasser85B, Gasser85C},  
calculations limited to the one loop order. In \cite{Coon86} the $\pi$-$\eta$ mixing matrix 
element is calculated from the decays of $\eta$ and $\eta^\prime$ considering $q^2$-independent 
$\pi$-$\eta$ vertex. 

At the hadronic level, $n$-$p$ mass difference $(M_n \neq M_p)$ causes to mix various isospin 
states in vacuum \cite{ Coon87, Blunden87, Langacker79, McNamee75, Coon77, Machleidt01, 
Coon87PRC26, Piekarewicz93PRC48}. The mixing amplitudes then 
used to construct the $CSB$ two body potential. In \cite{Blunden87} and \cite{Coon87} $CSB$ potentials 
have been constructed using on-shell and constant $\rho$-$\omega$ mixing amplitudes. Various $CSB$ observables 
have been calculated considering either constant or on-shell mixing amplitude and claimed successful 
\cite{Coon87, Machleidt01} for explaining the $CSB$ observables. Though the mixing amplitude 
as shown in \cite{Goldman92, Krein93, Oconnell94, Hutsuda94} has strong momentum dependence. On the 
basis of it the success of \cite{Coon87, Machleidt01} has been put into question \cite{Piekarewicz93,Cohen95}.

There is another class of mixing mechanism which is completely different in origin. The mixing of different 
mesons in $NM$ stems from the absorption and emission of intermediate mesons by neutron and proton Fermi spheres. 
Such mixing takes place in matter if the ground state contains unequal number of $n$ and $p$, {\em i.e.} 
$N \neq Z$ where $N$ and $Z$ represent the neutron and proton numbers respectively, called the asymmetric 
nuclear matter $(ANM)$. The asymmetry parameter $\alpha$ is defined as $\alpha= (\rho_n -\rho_p)/(\rho_n + \rho_p)$
where $\rho_n$ and $\rho_p$ correspond to the neutron and proton densities, (and $k_n$ and $k_p$ Fermi momenta). 

In $ANM$ the emission and absorption of different mesons by the neutron and proton Fermi spheres are such that 
their contributions do not cancel and it gives rise to a non vanishing mixing amplitude. If $N=Z$ {\em i.e.}
symmetric nuclear matter $(SNM)$, the contributions of neutron and proton Fermi spheres will cancel only 
if the ground state respects the symmetry. Otherwise, such cancellation does not take place even in $SNM$. 
Therefore, the matter induced mixing is an additional source of $CSB$ in $NN$ interaction.

The possibility of such matter induced mixing was first investigated by Dutt-Mazumder, Dutt-Roy and 
Kundu \cite{Abhee97} in ANM in the Walecka model \cite{Serot86} and subsequently in 
\cite{Broniowski98, Abhee01, Kampfer04, Roy08, Biswas06} similar investigations have been made. 
In \cite{Broniowski98} matter induced mixing was studied using phenomenological parametrization 
to incorporate the results of all models of meson properties in medium. Most of the above works 
investigated the role of matter induced mixing on dilepton spectrum, pion form factors, etc. 
Mori and Saito studied the properties of $CSB$ meson mixing in $ANM$ within the framework of 
quantum hadrodynamics and constructed $CSB$ potentials in spacelike region \cite{Saito03}. 
Similarly matter induced $\rho$-$\omega$ and $\pi$-$\eta$ mixing in $ANM$ have been investigated in 
\cite{Biswas:jan10, Biswas:jun10} and also constructed $CSB$ two body $NN$ potentials. 
In \cite{Saito03, Biswas:jan10, Biswas:jun10} the effect of asymmetry and medium to the $CSB$ potentials 
are incorporated through the in-medium mixing amplitudes. The mixing amplitudes are calculated using 
in-medium nucleon propagators \cite{Abhee97} which contains the medium modified nucleon mass and density 
of the $NM$. 

The main motivation of the present work is to study the medium effect, particularly 
the role of the in-medium $\pi$ and $\eta$ meson propagators, and effective nucleon 
mass to the $CSB$ potential. Inclusion of such in-medium meson propagators and medium
modified nucleon mass in the nucleon spinor of the external nucleon legs are consistent 
for the construction of $CSB$ potential in the nuclear medium which, in the previous works 
\cite{Saito03, Biswas:jan10, Biswas:jun10} were not considered. In addition, we also studied 
the role of effective masses of $\pi$ and $\eta$ mesons to the $CSB$ potential simply 
replacing their bare masses of the respective propagators .

The paper is organized as follows. We calculate $\pi$ and $\eta$ meson self-energies
in Sec.~\ref{sec:selfen:and:prop} considering both $PS$ and $PV$ interactions. 
These self-energies are then used to calculate the in-medium $\pi$ and $\eta$ meson propagators 
in sub section.~\ref{meson:propagators}. In Sec.~\ref{mixing:amplitude}
we calculate $\pi$-$\eta$ mixing amplitudes and construct the $CSB$ two body $NN$ class $III$ 
potentials in Sec.~\ref{csb:potential}. The Sec.~\ref{result} is devoted for presenting 
numerical results and discussion. And finally we summarize in Sec.~\ref{summary}.

\section{$\pi$ and $\eta$ meson self-energies and in-medium propagators}
\label{sec:selfen:and:prop}
The meson self-energies and propagators in the medium are essential ingredients of the present work. 
To calculate the meson self-energies in medium one uses the in-medium nucleon propagator $\gn_N(k)$ which 
consists of the Dirac sea and Fermi sea contributions known as the usual vacuum part $G_v(k)$, 
and the density dependent part $G_m(k)$, respectively \cite{Serot86}: 
\be
\gn_N(k)=G_v(k)+G_m(k)~, \label{nuc:prop}
\ee
where
\bse
\bea
G_v(k)\!\!&=&\!\!\frac{\sl{k}+\mn_N}{k^2-{\mn}^2_N+i\zeta}~, 
\label{vac:prop} \\
G_m(k)\!\!&=&\!\!\frac{i\pi}{\en_N}(\sl{k}+\mn_N)\delta(k_0-\en_N)
\theta(k_N-|\vk|)~.
\label{med:prop}
\eea
\ese
\indent In the above equations $N$ represents the nucleon index $p$ for proton and $n$ for neutron and 
$k=(k_0,{\bf k})$ denotes the four momentum of the loop nucleon. The nucleon energy is denoted by 
$\en_N=\sqrt{{\mn}^2_N+{\bf k}^2}$, where $\widetilde{M}_N$ is the in-medium nucleon mass. 

Note that the $\delta$-function in Eq.~(\ref{med:prop}) indicates the nucleons are on-shell and the 
$\theta$-function, called Pauli blocking, ensures that the momentum of the nucleon propagating 
in the medium must be less than the Fermi momentum $k_N$. 

In quantum hadrodynamics $(QHD-I)$ the nucleon is assumed to move in the mean field produced by 
the neutral scalar and vector mesons \cite{Serot86, Walecka74, Chin74, Chin77, Boguta:NPA77, Boguta:PLB77, 
Serr78}. The scalar field modifies the nucleon mass as,
 
\be
\mn_N=M_N-\frac{\rm g^2_\sigma}{m^2_\sigma}(\rho^s_p+\rho^s_n) ~.\label{nm}
\ee
while the vector field causes to shift the energy which we do not consider in this work. 

In the above equation $m_\sigma$ and $g_\sigma$ represent the mass and coupling constant 
of the scalar meson $\sigma$, respectively. The bare nucleon mass is denoted by, $M_N$ and 
$\rho^s_N$ denotes the scalar density:
\be
\rho^s_N=\frac{\mn_N}{2\pi^2}\left[\en_Nk_N-\mn^2_N \ln\left(\frac{\en_N+k_N}{\mn_N} \right)\right]~.
\label{scal:dens}
\ee
The in-medium nucleon mass $\widetilde{M}_N$ can be determined from Eq.~(\ref{nm}) solving 
it self consistently.

\subsection{The Meson Self-Energies}
\label{meson:selfenergy}
\begin{figure}[htb!]
\begin{center}
\includegraphics[scale=0.45,angle=0]{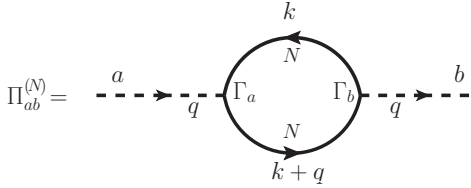}
\caption{The generic diagram of meson self-energy at the one loop order.}
\label{fig:selfen01}
\end{center}
\end{figure}

The calculations of the in-medium meson self-energies have been restricted up to one loop order
(see Fig.~\ref{fig:selfen01}) which have been calculated using the formula \cite{Peskin95}:
\bea
\Pi^{(\!N\!)}_{ab}\!(q^2)\!\!&\!=&\!\!-i\!\!\!\int\!\!\frac{d^4k}{(2\pi)^4}
\mathrm{Tr}\!\left[\Gamma_a(q) \gn_N(k) \Gamma_b(-q) \gn_N(k\!+\!q)\right]~,\nn \\
\label{selfen01}
\eea
\noindent where $a~({\rm or}~ b)$ denotes $\pi$ or $\eta$ and $\Gamma_{a,b}(q)$ represent the 
meson-nucleon-nucleon vertex factor. The dashed lines in Fig.~\ref{fig:selfen01} represent the 
meson propagators and the solid lines denote the in-medium nucleon propagators. 

Substituting the in-medium nucleon propagator, $G_N(k)$ in the expression of 
Eq.~(\ref{selfen01}) one may distinguish four terms with the combinations like 
${\Gamma_a}G_v{\Gamma_b}G_v$, ${\Gamma_a}G_v{\Gamma_b}G_m$, ${\Gamma_a}G_m{\Gamma_b}G_v$ 
and ${\Gamma_a}G_m{\Gamma_b}G_m$. Out of which first three terms have been considered  
in this paper for calculating the total meson self-energy in the medium. The fourth term contains two 
$\delta$-functions which means both the loop nucleons, in Fig.~\ref{fig:selfen01}  are on-shell which 
means that the mesons will decay into nucleon - antinucleon pair and it happens if the meson momentum 
is larger than two times the nucleon momentum, {\em i.e.} $q>2k_N$ (also $q_0>2\en_N$). Since the 
present work is restricted only for the low energy collective excitation of mesons in the medium, 
the fourth term has been neglected \cite{Chin77}. 

The meson self-energy is the sum of the contributions of $p$ and $n$ loops: 
\be
\Pi_{aa}(q^2)=\Pi^{(p)}_{aa}(q^2)+\Pi^{(n)}_{aa}(q^2).
\ee 

Note that the part of the self-energy containing the term ${\Gamma_a}G_v{\Gamma_b}G_v$ is called 
vacuum part while the other part {\em i.e.} the sum of the terms ${\Gamma_a}G_v{\Gamma_b}G_m$
and ${\Gamma_a}G_m{\Gamma_b}G_v$ is called the medium part of the self-energy. 

\subsubsection*{{\bf A1. Pseudoscalar interaction}}
We shall first calculate the self-energies of $\pio$ and $\eta$ mesons considering the pseudoscalar 
interactions described by the Lagrangians 
\bse
\label{ps:int}
\bea
{\cal L}^{PS}_{\pio}&=&-i\g_\pi \left[ \bar{\Psi}_p{\gamma_5}\Phi_{\pio}\Psi_p -
\bar{\Psi}_n{\gamma_5}\Phi_{\pio}\Psi_n \right]~, \label{ps:pi}\\
{\cal L}^{PS}_{\eta}&=&-i\g_\eta \left[ \bar{\Psi}_p{\gamma_5}\Phi_{\eta}\Psi_p +
\bar{\Psi}_n{\gamma_5}\Phi_{\eta}\Psi_n \right]~. \label{ps:eta}
\eea
\ese

In the above Lagrangians, $\Psi$ and $\phi$ represent the nucleon spinor and meson fields,
respectively. $\g_\pi~(\g_\eta)$ is the meson-nucleon coupling constants. The vertex factor 
$\Gamma_a(q)=-i\g_a\gamma_5$. 
The contribution of nucleon loop  ({\em i.e.} either proton or neutron loop) at one 
loop order to the vacuum part of the meson self-energy can be written from Eq.~(\ref{selfen01}) as
\bla
&\Pi^{(\!N\!)}_{ab,v}(\!q^2\!)\!\!\!=\!-i\!\!\int\!\!\frac{d^4k}{(2\pi)^4}
\mathrm{Tr}\!\left[(\!-i\g_a\gamma_5) G_v(k)(\!-i\g_b\gamma_5)G_v(k\!+\!q)\right] && \nn \\
&=\!4i\g_a\g_b\!\int\! \frac{d^4k}{(2\pi)^4}
\left[\frac{\mn^2_N-k\cdot(k+q)}{(k^2-\mn^2_N)((k+q)^2-\mn^2_N)}\!\right]~. &&
\label{ps:vac:selfen01}
\ela

The above self-energy integral is found to be quadratically divergent at the one loop order 
and needs to be regularized. The divergent terms can be isolated following 
the method of dimensional regularization \cite{Hooft73,Peskin95,Cheng06}. The $PS$ interaction 
is renormalizable by adding  appropriate counter terms to the Lagrangian \cite{Matsui82,Serot86}. 
Renormalizable means the self-energy integral becomes divergence free for all orders \cite{Mornas02}.
The renormalized vacuum contribution to the self-energy of nucleon loop has been borrowed from
Ref.~\cite{Biswas08}:
\bla
&\Pi^{(\!N\!)}_{aa,v}(\!q^2\!)
\!\!=\!\!\left(\!\frac{\g_a}{2\pi}\!\!\right)^2
\!\!\left[3(\mn^2_N-M^2_N)+(q^2-m^2_a) \left(\frac{1}{6}+\frac{M^2_N}{m^2_a}\right)\right. && \nn \\ 
&-2\mn^2_N\ln\left(\frac{\mn_N}{M_N}\right)+\frac{8M^2_N(M_N-\mn_N)^2}{(4M_N^2-m^2_a)}&& \nn \\
&-\frac{2\mn^2_N\sqrt{4\mn^2_N-q^2}}{q} \tan^{-1}\left(\frac{q}{\sqrt{4\mn^2_N-q^2}} \right) &&\nn\\
&+\frac{2M^2_N\sqrt{4M^2_N-m^2_a}}{m_a}\tan^{-1}\left(\frac{m_a}{\sqrt{4M^2_N-m^2_a}} \right)
&&\nn \\
&+ \left( (M^2_N-\mn^2_N)+\frac{m^2_a(M_N-\mn_N)^2}{(4M^2_N-m^2_a)}
+\frac{M^2_N(q^2-m^2_a)}{m^2_a} \right) &&\nn \\
&\times  \frac{8M^2_N}{m_a\sqrt{4M^2_N-m^2_a}} 
\tan^{-1}\left(\frac{m_a}{\sqrt{4M^2_N-m^2_a} }\right)&&\nn \\
&+\left. \int^1_0 dx~3x(1-x)q^2  \ln\left(\frac{\mn^2_N-q^2x(1-x)}{M^2_N-m^2_a x(1-x)} \right)\right].&&
\label{ps:vac:selfen02}
\ela
The vacuum part of the meson self-energy may be written after suitable approximation as
\bea
\Pi^v_{aa}(q^2)&\approx & A_{0a} + A_{1a} q^2~, \label{ps:vac:selfen03}
\eea
keeping terms up to orders $q^2/M^2_N$ and neglecting its higher orders, where 
\bse
\bea
A_{0a}\!\!&=&\!\!\left(\!\g_a\over{2\pi}\!\!\right)^2 
\!\!\left[3(\mn_p^2 -2M_p^2)+2\mn^2_p\ln\left(\frac{M_p}{\mn_p}\right)\right. \nn \\ 
&\!\!+\!\!&\left.3(\mn_n^2-2M_n^2)+2\mn^2_n\ln\left(\frac{M_n}{\mn_n}\right)\right], 
\label{ps:vac:a0}\\
A_{1a}\!\!&=&3\left(g_a\over{2\pi}\right)^2(M^2_n+M^2_p)/{m^2_a}~.\label{ps:vac:a1}
\eea
\ese
The contribution of the medium part of the self-energy reads
\bea
\Pi^{(\!N\!)}_{ab,m}(\!q^2\!)\!\!\!&=&\!\!\!-i\!\!\int\!\!\frac{d^4k}{(2\pi)^4}
{\bf Tr}\left[(\!-i\g_a\gamma_5\!)G_v(k)(\!-i\g_b\gamma_5) G_m(k\!+\!q) \right.\nn\\
&+&\left.(-i\g_a\gamma_5) G_m(k)(-i\g_b\gamma_5)G_v(k\!+\!q)\right]~,
\eea
which after calculating the trace and performing the integration of $k_0$, can be written
as
\bea
\Pi^{(\!N\!)}_{ab,m}(\!q^2\!)\!\!\!&=&\!\!\!-2{\g_a}{\g_b}\!\!\int\!\!\frac{d^3k}{(2\pi)^3\!\en_N} 
\!\left[\!{4(k\cdot{q})^2\over{q^4\!-\!4(k.q)^2}}\!\right]\! \theta(k_N\!-\!|\vk|). \nn\\
&& \label{ps:med:seflen01}
\eea
The calculation is restricted to the low momentum excitation as mentioned earlier. 
That means $q^2<k^2$. This allows one to write Eq.~(\ref{ps:med:seflen01}) as 
\bea
\Pi^{(N)}_{ab,m}(\!q^2\!)\!\!&\approx&\!\!2{\g_a}{\g_b}\!\!\int\!\!\frac{d^3k}{(2\pi)^3\!\en_N} 
\left[\!1\!+\!{q^4\over{4(k.q)^2}}\!\right]\!\theta(k_N\!-\!|\vk|),\nn \\
&& \label{ps:med:selfen02}
\eea
which after integration and algebraic manipulation (see {\bf Appendix \ref{appndixa}}) reduces to

\bla
&\Pi^{(N)}_{ab,m}(q^2)\!\approx 2\left(\!{\g_a}\over{2\pi}\!\right)\!\!\left(\!{\g_b}\over{2\pi}\!\right)&&\nn \\
&\!\times\!\left[\!\left\{k_N\en_N\!-\!{1\over2}\!\mn^2_N\!\ln\left(\!\frac{\en_N\!+\!k_N}{\en_N\!-\!k_N}\!\right)
\!\!\right\}\right. && \nn \\
\!\!&+\!\!(q^2_0-2\vq^2){1\over4}
\!\!\!\left.\left\{\frac{k_N\en_N}{\mn^2_N}\!-\!{1\over2}\ln\left(\!\frac{\en_N\!+\!k_N}{\en_N\!-\!k_N}\!\right)
\!\!\right\}\!\!\right].&& \label{ps:med:selfen03}
\ela
Similar to the vacuum part one may approximate the medium part of the self-energy.  
\bea
\Pi_{aa,m}(\!q^2\!)&\approx&B_{0a}+B_{1a}(q^2_0-2\vq^2)~,\label{ps:med:selfen04}
\eea
\noindent where
\bse
\bea
B_{0a}\!\!\!&=&\!\!2\!\left(\g_a\over{2\pi}\!\right)^2\!\left[\!\left(k_p\en_p\!+\! k_n\en_n\!\right)\right. \nn \\
&-&\!\!{1\over2}\!\!\left.\left\{\!\mn^2_p\ln\left(\!\frac{\en_p\!+\!k_p}{\!\en_p\!-\!k_p}\!\right)
\!\!+\!\!\mn^2_n\ln\left(\!\frac{\en_n\!+\!k_n}{\en_n\!-\!k_n}\!\right)\!\right\}\!\right],~~\label{ps:med:b0} \\
B_{1a}\!\!\!&=&\!\!\!\left(\!\g_a\over{2\pi}\!\right)^2\!\! {1\over2} 
\left[\!\!\left(\!{k_p\en_p\over\mn^2_p}\!+\!{k_n\en_n\over\mn^2_n}\!\!\right)\right. \nn \\
&-&{1\over2}\left.\left\{\ln\left(\!\frac{\en_p\!+\!k_p}{\en_p\!-\!k_p}\!\right)
\!+\!\ln\left(\!\frac{\en_n\!+\!k_n}{\en_n\!-\!k_n}\!\right)\!\!\right\}\!\right]~.\label{ps:med:b1}
\eea
\ese
The total self-energy is the sum of vacuum and medium contributions: 
\be
\Pi_{aa}(q^2) = \Pi_{aa,v}(q^2) + \Pi_{aa,m}(q^2)~.
\ee
The space like self-energy is required for construction of the CSB potential in momentum space which
is obtained by substituting $q_0 = 0$ in the expression of $\Pi_{aa}$, 
\be
\Pi^{PS}_{aa}(\vq^2)= (A_{0a}+B_{0a})- (A_{1a}+2B_{1a})\vq^2 ~.\label{ps:total:selfen01}
\ee

\subsection*{{\bf A2. Pseudovector interaction}}
In this section we calculate the meson self-energies considering pseudovector meson-nucleon
interactions: 
\bse
\label{pv:int}
\bea
{\cal L}^{PV}_{\pio}\!\!\!&=&\!\!\!-
\gpim\!\!\left[\!\bar{\Psi}_p{\gamma_5}\sl{\partial}\Phi_{\pio}\!\Psi_p\!-\!
\bar{\Psi}_n{\gamma_5}\sl{\partial}\Phi_{\pio}\!\Psi_n\!\right]~,\label{pv:pi}\\
{\cal L}^{PV}_{\eta}\!\!\!&=&\!\!\!- 
\getam\!\!\left[\bar{\Psi}_p{\gamma_5}\sl{\partial}\Phi_{\eta}\Psi_p\!+\!
\bar{\Psi}_n{\gamma_5}\sl{\partial}\Phi_{\eta}\Psi_n \right]~.\label{pv:eta}
\eea
\ese
Following Eq.~(\ref{selfen01}), one may write the contribution of nucleon loop to the
vacuum part of the self-energy
\bla
&\Pi^{\prime(N)}_{ab,v}(q^2)=-i\int\frac{d^4k}{(2\pi)^4} &&  \nn \\
&\times\mathrm{Tr}\left[\left(-\gam\gamma_5\sl{q}\right) G_v(k)\left(\gbm\gamma_5\sl{q}\right)
G_v(k\!+\!q)\right]~, && \nn \\
&=4i\left(\gam\right) \left(\gbm\right)\int \frac{d^4k}{(2\pi)^4} && \nn \\
&\times\left[\frac{q^2(\mn^2_N-k\cdot(k+q))-2(k\cdot q)q\cdot(k+q) }{(k^2-\mn^2_N)( (k+q)^2-\mn^2_N)} 
\right]~.&&\label{pv:vac:selfen00}
\ela

The above integral is also divergent. Similar to the $PS$ interaction one may invoke dimensional 
regularization to extract the diverging parts of this integral. After dimensional regularization 
the integral of Eq.~(\ref{pv:vac:selfen00}) reduces to
\bla
&\Pi^{\prime(N)}_{ab,v}(q^2)=\frac{1}{2}\left(\frac{\mathrm{g}_a}{2\pi} \right)
\left(\frac{\mathrm{g}_b}{2\pi} \right)q^2 &&\nn \\
&\times \left(\frac{\mn_N}{M_N}\right)^2\!\!\left[- 2 -\frac{1}{\epsilon}+\gamma_E+\ln(4\pi\mu^2)
+2\ln(\mn_N)  \right. &&\nn\\ 
&+\left.2\frac{\sqrt{4\mn^2_N-q^2}}{q}\tan^{-1}\left(\frac{q}{\sqrt{4\mn^2_N-q^2}} \right) \right] &&
\label{pv:vac:selfen01}
\ela
Note that $\epsilon = (4 - D)/2$ and $\mu$ is an arbitrary scaling parameter. $D$ is the dimension of
integration. $\gamma_E$ is the Euler-Mascheroni constant. The above integral diverges for 
$D=4$. 

The pseudovector interaction is non-renormalizable because of the derivative term. That means
one can not eliminate the divergences for all orders by adding appropriate counterterms in the Lagrangian
\cite{Mornas02}. Various renormalization methods have been discussed in \cite{Mornas02}. It 
is to be mentioned that the result depends on a particular method \cite{Mornas02}.  Here we adopt 
subtraction scheme \cite{Biswas08} to eliminate the divergences ({\em see}~{\bf Appendix} 
\ref{appendixb}) of Eq.~(\ref{pv:vac:selfen01}): 
\bla
&\Pi^{\prime(N)}_{aa,v}(q^2)=\left(\frac{\mathrm{g}_a}{2\pi} \right)^2q^2&&\nn \\
&\times\left(\frac{\mn_N}{M_N}\right)^2
\left[\frac{\sqrt{4\mn^2_N-q^2}}{q}\tan^{-1}\left(\frac{q}{\sqrt{4\mn^2_N-q^2}} \right)\right.&&\nn\\
&+\left.\frac{\sqrt{4\mn^2_N-m^2_a}}{m_a}\tan^{-1}\left(\frac{m_a}{\sqrt{4\mn^2_N-m^2_a}} \right) \right]~.&& 
\label{pv:vac:selfen02}
\ela

Now one may approximate vacuum part of the self-energy similar to the $PS$ interaction: 
\bea
\Pi^{\prime}_{aa,v}(q^2)&\approx& A^{\prime}_{1a}q^2~, \label{pv:vac:selfen03}
\eea
where
\be
A^{\prime}_{1a}=\frac{1}{12}\left(\frac{\mathrm{g}_a}{2\pi}\right)^2 
\left(\frac{m^2_a}{M^2_p}+\frac{m^2_a}{M^2_n} \right)~. \label{pv:vac:a1}
\ee

The contribution of nucleon loop to the medium part of the meson self-energy reads 
\bla
&\Pi^{\prime(N)}_{ab,m}(q^2)\!\!=\!\!-i\!\!\int\!\!\frac{d^4k}{(2\pi)^4}&&\nn \\
\!\!\!&\times\!\!\mathrm{Tr}\left[\left(-\gam\gamma_5\sl{q}\right)G_v(k)
\left(\gbm\gamma_5\sl{q}\right)G_m(k\!+\!q)\right.&&\nn\\
\!\!&+\!\!\left.\left(-\gbm\gamma_5\sl{q}\right)G_m(k)
\left(\gam\gamma_5\sl{q}\right)G_v(k\!+\!q)\right]~,&&
\ela

which after calculating the trace and performing the integration of $k_0$ similar to that of 
the medium part of $PS$ interaction, reduces to
\bea
\Pi^{\prime(N)}_{ab,m}(q^2)\!\!&=&\!\!\!-8\!\!\left(\!\!\gam\!\!\right)\!\!\left(\!\!\gbm\!\!\right)\mn^2_Nq^4 \nn \\
\!\!\!&\times\!\!\!&\!\!\!\int\!\!\frac{d^3k}{(2\pi)^3\!\en_N}\!\left[\frac{1}{q^4\!-\!4(k.q)^2}\right]\!
\theta(k_N\!-\!|\vk|)~.~ \label{pv:med:selfen01}
\eea

Now the Eq.~(\ref{pv:med:selfen01}) may be evaluated and approximated as discussed in 
{\bf Appendix} \ref{appndixa}:
\bla
&\Pi^{\prime(N)}_{ab,m}(q^2)\approx (q^2_0-2\vq^2)\left(\frac{\mathrm{g}_a}{2\pi}\right)
\left(\frac{\mathrm{g}_b}{2\pi} \right)&&  \nn \\
&\times\frac{1}{2}\left[\frac{k_N\en_N}{M^2_N}-\frac{1}{2} \left(\frac{\mn_N}{M_N}\right)^2
\ln\left(\frac{\en_N+k_N}{\en_N-k_N}\right)\right]~.&&
\label{pv:med:selfen02}         
\ela

The medium part of the self-energy can be written as
\bea
\Pi^{\prime(N)}_{aa,m}(q^2)&\approx& B^\prime_{1a}(q^2_0-2\vq^2) \label{pv:med:selfen03}
\eea
where, 
\bea
B^\prime_{1a}&=& \left(\frac{\mathrm{g}_a}{2\pi} \right)^2 
\frac{1}{2}\left[\left(\frac{k_p\en_p}{M^2_p}+\frac{k_n\en_n}{M^2_n}\right) \right. \nn \\
&-&{1\over2} \left\{ \left(\frac{\mn_p}{M_p} \right)^2\ln\left(\frac{\en_p+k_p}{\en_p-k_p} \right)\right. \nn \\
&+&\left.\left.\left(\frac{\mn_n}{M_n} \right)^2\ln\left(\frac{\en_n+k_n}{\en_n-k_n} \right) \right\} \right]~.
\label{pv:med:b1}
\eea
The total space like self-energy is given by
\be
\Pi^{\prime PV}_{aa} = - (A^{\prime}_{1a}  + 2 B^{\prime}_{1a} )\vq^2~. \label{pv:total:selfen01}
\ee
\subsection{Meson Propagators in medium}
\label{meson:propagators}
While mesons propagate through the nuclear medium they are scattered by the Fermi spheres 
and receives corrections to their masses and energies. This effect may be incorporated through 
the in-medium meson propagators. Such in-medium propagator may be derived from standard 
covariant perturbation theory. Here we will solve the Swinger - Dyson equation to find the in-medium 
meson propagator:
\be
\widetilde{D}_{a}(q^2) = D_{a}(q^2) + D_{a}(q^2)\Pi_{aa}(q^2)D_{a}(q^2) \cdots + \cdots 
\label{eqn:dyson01}
\ee
where $D_{a}(q^2)=(q^2 - m^2_{a}+i\epsilon)^{-1}$ is representing the bare meson propagator while
$\widetilde{D_a}(q^2)$ is the in-medium meson propagator which after solving Eq.~(\ref{eqn:dyson01})
can be written as 
\be
\widetilde{D}_{a}(q^2)^{-1} = q^2 - m^2_{a} -\Pi_{aa}(q^2)~. \label{eqn:dyson02}
\ee

\begin{figure}[htb!]
\begin{center}
\includegraphics[scale=0.35,angle=0]{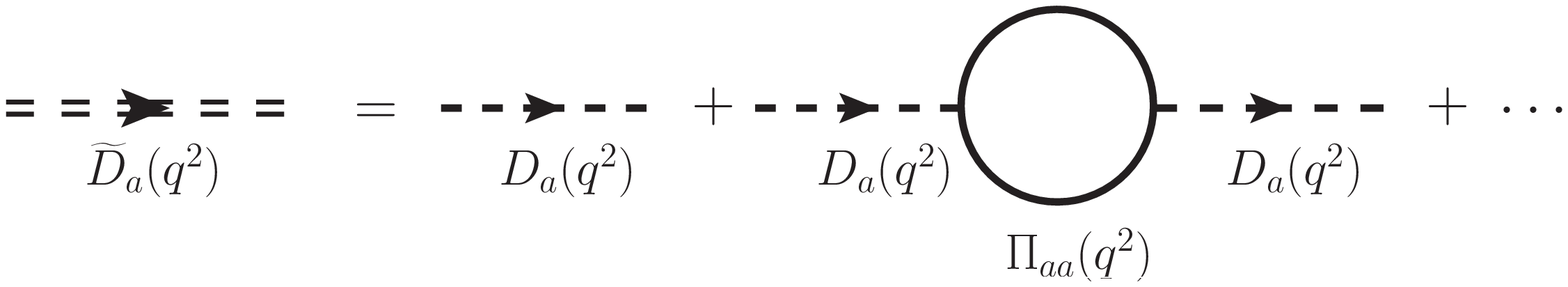}
\caption{Diagrammatic presentation of Swinger - Dyson equation.}
\label{medprop}
\end{center}
\end{figure}

In Eq.~(\ref{eqn:dyson02}) the imaginary part, $i\epsilon$ has been dropped as it is not important in 
the present context. 

For the construction of CSB potential one needs space like in-medium meson propagators which 
is obtained from Eq.~(\ref{eqn:dyson02}) by substituting $q_0=0$ in this equation. The total 
meson self-energies from Eq.~(\ref{ps:total:selfen01}) and Eq.~(\ref{pv:total:selfen01}) in 
Eq.~(\ref{eqn:dyson02}) one can write the space like meson propagators $\widetilde{D}_a(\vq^2)$ 
and $\widetilde{D}^{\prime}_a(\vq^2)$ in medium for $PS$ and $PV$ interactions respectively,
\bse
\bea
\widetilde{D}_a(\vq^2) & =&  \frac{-1}{(1-A_{1a}-2B_{1a})(\vq^2+M^2_a)} \label{ps:med:prop}\\
\widetilde{D}^{\prime}_a(\vq^2) & =& \frac{-1}{(1-A^{\prime}_{1a}-2B^{\prime}_{1a})(\vq^2+M^{\prime 2}_a)}
\label{pv:med:prop}
\eea
\ese
where 
\bse
\bea
M^2_a &=& \frac{m^2_a+A_{0a}+B_{0a}}{1-A_{1a}-2B_{1a}} \\
M^{\prime2}_a &=& \frac{m^2_a}{1-A^{\prime}_{1a}-2B^{\prime}_{1a}}
\eea
\ese
Note that $M_a$ ( or  $M^\prime_a$) is not the effective mass of meson. The effective mass of meson
can be obtained from Eq.~(\ref{eqn:dyson02}) by solving $\widetilde{D}_{a}(\vq^2=0)^{-1}=0$. 

\section{$\pi$-$\eta$ mixing amplitude}
\label{mixing:amplitude}
\begin{figure}[htb!]
\begin{center}
\includegraphics[scale=0.35,angle=0]{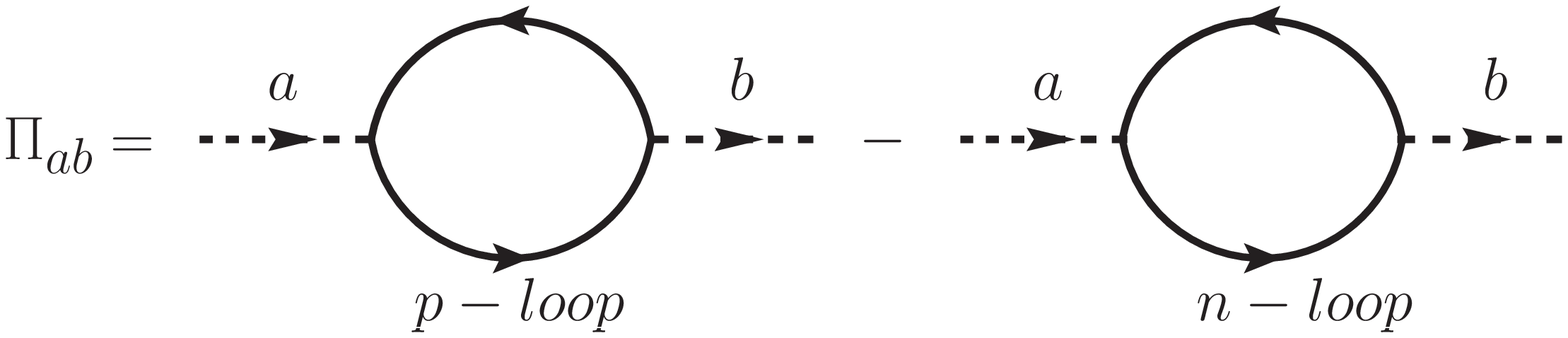}
\caption{Generic diagram for mixing amplitude.}
\label{mixing}
\end{center}
\end{figure}

In this section $\pi$-$\eta$ mixing amplitudes have been calculated at the one loop order 
for both $PS$ and $PV$ interactions. The mixing amplitude, $\Pi_{ab}(q^2)$ is generated by the 
$p$-loop contribution, $\Pi^{(p)}_{ab}(q^2)$ minus the $n$-loop contribution, $\Pi^{(n)}_{ab}(q^2)$ 
as shown in Fig.~\ref{mixing} where, the continuous line represents the loop nucleon and mesons 
by the dashed lines. \\
\vspace{-0.25cm}\\
The origin of negative sign between the $p$-loop and $n$-loop contributions may be understood 
from the interaction Lagrangians given in Eq.~(\ref{ps:int}) or Eq.~(\ref{pv:int}). Note that $\pi^0$
and $\eta$ couples to proton with the same sign while they couple to the neutron with opposite 
sign. This brings the negative sign between $p$ and $n$ loop contributions. Therefore the mixing 
amplitude, 
\be
\Pi_{ab}(q^2) = \Pi^{(p)}_{ab}(q^2) - \Pi^{(n)}_{ab}(q^2)~.
\ee

First we proceed to calculate the mixing amplitude for $PS$ interaction. 
\vspace{-0.45cm}
\subsection{Pseudoscalar interaction}
The vacuum contribution of the nucleon loop to the $\pi$-$\eta$ mixing amplitude can be obtained 
from Eq.~(\ref{ps:vac:selfen01}). As discussed before, this part contains divergent terms, one 
of  which is proportional to $\mn_N$. This divergent term may be eliminated by subtracting 
$\Pi^{(N)}_{\pi\eta,v}(q^2=0)$ from $\Pi^{(N)}_{\pi\eta,v}(q^2)$. After this subtraction the 
vacuum contribution reads
\bla
&\Pi^{(\!N\!)}_{\pi\eta,v}\!(q^2)\!=\!\left(\!{\g_\pi\over{2\pi}}\!\right)\!\!\left(\!{\g_\eta\over{2\pi}}\!\right) q^2
\left[\!{1\over{2}}\!\!\left(\!\!-{7\over3}-{1\over\epsilon}+ \gamma_E -\ln(4\pi\mu^2)\right)\right.&&\nn \\
&\!\!+\left.\!\!\ln(\mn_N)\!\!+\!\frac{\sqrt{4\mn^2_N-q^2}}{q}
\tan^{-1}\left(\!\!\frac{q}{\sqrt{4\mn^2_N\!-\!q^2}}\!\!\right)\!\right]~.&& \label{ps:vac:mix01}
\ela

Note that this simple subtraction, however, removes the divergence proportional 
to $\mn^2_N$ but can not remove the divergence completely. There is still one divergent term proportional 
to $q^2$ as seen in Eq.~(\ref{ps:vac:mix01}). The subtraction of $n$-loop contribution from the 
$p$-loop contribution completely removes this divergent term yielding the vacuum part of the 
mixing amplitude finite. Thus the vacuum part of the mixing amplitude can be approximated as
\bea
\Pi^{PS}_{\pi\eta,v}(q^2)&\approx& A_{1\pi\eta}q^2~, \label{ps:vac:mix02}
\eea
\noindent where the constant
\be
A_{1\pi\eta}=\left({\g_\pi\over{2\pi}} \right)\left({\g_\eta\over{2\pi}} \right)\ln(\mn_p/\mn_n)~.
\label{ps:vac:mix:a1}
\ee

It is important to note that in case of mixing in vacuum, the mixing amplitude can be obtained
by replacing $\mn_p ({\rm or}~\mn_n)$ with $M_p ({\rm or}~M_n)$ in Eq.~(\ref{ps:vac:mix:a1}). 
The mixing amplitude vanishes if  $M_p=M_n$ and CSB interaction in vacuum vanishes. 

The medium contribution to the mixing amplitude can be obtained from the Eq.~(\ref{ps:med:selfen03}). 
Therefore the medium part of the mixing amplitude may be written as
\bea
\Pi^{PS}_{\pi\eta,m}(q^2)&\approx& B_{0\pi\eta} + B_{1\pi\eta}(q^2_0-2\vq^2)~.\label{ps:med:mix01}
\eea
\noindent The constants
\bse
\bea
B_{0\pi\eta}\!\!\!&=&\!\!\!2\left(\g_\pi\over{2\pi}\right)\!\!\left(\g_\eta\over{2\pi}\right)\!\!
\left[\!\left(\!k_p\en_p- k_n\en_n\!\right)\right.\nn \\
&\!\!-&\left.\!\!{1\over2}\!\left\{\!\!\mn^2_p\ln\left(\!\!\frac{\en_p\!+\!k_p}{\en_p\!-\!k_p}\!\!\right)
\!\!-\!\!\mn^2_n\ln\left(\!\!\frac{\en_n\!+\!k_n}{\en_n\!-\!k_n}\!\!\right)\!\!\right\}\!\!\right]~,
\label{ps:med:mix:b0}\\
B_{1\pi\eta}\!\!\!&=&\!\!\!\left(\g_\pi\over{2\pi}\right)\!\!\left(\g_\eta\over{2\pi}\right)
\!\!{1\over2}\left[\!\left(\!\!{k_p\en_p\over\mn^2_p}\!-\!{k_n\en_n\over\mn^2_n}\!\!\right)\right.\nn \\
&\!\!-&\left.\!\!{1\over2}\!\!\left\{\ln\!\left(\!\!\frac{\en_p\!+\!k_p}{\en_p\!-\!k_p}\!\!\right)
-\!\ln\!\left(\!\!\frac{\en_n\!+\!k_n}{\en_n\!-\!k_n}\!\!\right)\!\!\right\}\!\!\right]~.\label{ps:med:mix:b1}
\eea
\ese

The in-medium nucleon mass $\mn_N$ and nucleon energy $\en_N$ depend on the Fermi momentum 
$k_N$ of the nucleon, which is a function of baryon density $\rho_B$ and the asymmetry parameter 
$\alpha$ as discussed in section \ref{sec:selfen:and:prop}. Therefore the constants, $A_{1\pi\eta}$,
$B_{0\pi\eta}$ and $B_{1\pi\eta}$ depend on $\rho_B$ and $\alpha$. Thus for the mixing amplitudes 
in this case, both the vacuum and medium parts are driven by the asymmetry parameter $\alpha$.\\

To construct $CSB$ potential in momentum space one needs space like mixing amplitude which
reads as
\bse
\bea
\Pi^{PS}_{\pi\eta,v}(\vq^2)&\approx& - A_{1\pi\eta}\vq^2~, \label{ps:vac:mix03} \\
\Pi^{PS}_{\pi\eta,m}(\vq^2)&\approx& B_{0\pi\eta}-2 B_{1\pi\eta}\vq^2~.\label{ps:med:mix02}
\eea
\ese
The mixing amplitude contains both the vacuum and medium parts,  
\be
\Pi^{PS}_{\pi\eta}(\vq^2)=\Pi^{PS}_{\pi\eta,v}(\vq^2)+\Pi^{PS}_{\pi\eta,m}(\vq^2)~.
\ee

\subsection{Pseudovector interaction}
The vacuum contribution of $\pi$-$\eta$ mixing amplitude in case of $PV$ interaction can be 
obtained from Eq.~(\ref{pv:vac:selfen02}) following the method discussed in {\bf appendix \ref{appendixb}}.
The vacuum contribution of the mixing amplitude therefore reads
\bea
\Pi^{\prime PV}_{\pi\eta,v}(q^2)&\approx& A^\prime_{1\pi\eta}q^2~. \label{pv:vac:mix01}
\eea 

Note that the value of the constant $A^\prime_{1\pi\eta}$ depends on the condition (\ref{appendixb:vac:02}).
If one consider $m^2_a=m^2_\pi$ in Eq.~(\ref{appendixb:vac:02}) then 
\be
A^\prime_{1\pi\eta}\!\!=\!\!\frac{1}{12}\left(\frac{\mathrm{g}_a}{2\pi}\right)^2 
\left(\frac{m^2_\pi}{M^2_p}-\frac{m^2_\pi}{M^2_n} \right)~,\label{pv:vac:mix:a1}
\ee
and if $m^2_a=m^2_\eta$ is chosen in Eq.~(\ref{appendixb:vac:02}) then 
\be
A^\prime_{1\pi\eta}\!\!=\!\!\frac{1}{12}\left(\frac{\mathrm{g}_a}{2\pi}\right)^2 
\left(\frac{m^2_\eta}{M^2_p}-\frac{m^2_\eta}{M^2_n} \right)~.\label{pv:vac:mix:a2}
\ee

The vacuum contribution to the mixing amplitudes as obtained above also vanish in the limit $M_p=M_n$, 
but unlike to the $PS$ interaction, independent of medium effect because of the method adopted
to remove the divergence. The density dependent part of the mixing amplitude may be obtained from 
Eq.~(\ref{pv:med:selfen02}):

\bea
\Pi^{\prime PV}_{\pi\eta, m}(q^2)&\approx& B^\prime_{1\pi\eta}(q^2_0-2\vq^2)~, \label{pv:med:mix01}
\eea 
where the constant $B^\prime_{1\pi\eta}$ is given by
\bla
&B^\prime_{1\pi\eta}\!=\!\left(\frac{\g_\pi}{2\pi}\!\right)\!\!\left(\!\frac{\g_\eta}{2\pi}\!\right) 
\!\frac{1}{2}\!\left[\!\!\left(\!\!\frac{k_p\en_p}{M^2_p}\!-\!\frac{k_n\en_n}{M^2_n}\!\!\right)\right. &&\nn \\
&\!\!-\!{1\over2}\left.\!\!\left\{\!\!\left(\!\frac{\mn_p}{M_p}\!\right)^2
\!\!\!\ln\!\left(\!\!\frac{\en_p\!+\!k_p}{\en_p\!-\!k_p}\!\!\right)\! 
-\left(\!\frac{\mn_n}{M_n}\!\right)^2
\!\!\!\ln\!\left(\!\!\frac{\en_n\!+\!k_n}{\en_n\!-\!k_n}\!\!\right)\!\!\right\}\!\!\right].&&
\label{pv:med:mix:b1}
\ela
The space like mixing amplitudes are given by
\bse
\bea
\Pi^{\prime PV}_{\pi\eta,v}(\vq^2)&\approx& - A^\prime_{1\pi\eta}\vq^2~, \label{pv:vac:mix02}\\
\Pi^{\prime PV}_{\pi\eta,m}(\vq^2)&\approx&-2 B^\prime_{1\pi\eta}\vq^2~. \label{pv:med:mix02}
\eea
\ese
\section{Charge symmetry breaking potential}
\label{csb:potential}
\begin{figure}[hbt!]
\begin{center}
\includegraphics[scale=0.285,angle=0]{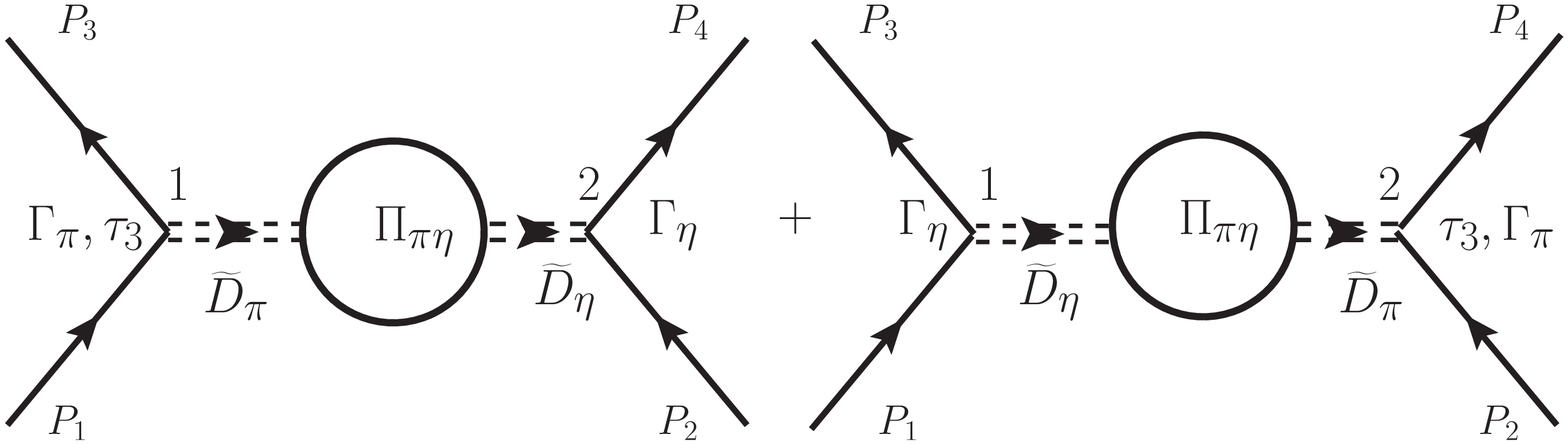}
\caption{Diagrams contribute to $CSB$ potential. The solid straight lines and double dashed lines 
represent external nucleons and the in-medium meson propagators, respectively. The $CSB$ amplitudes
are indicated by the circles.}
\label{csb}
\end{center}
\end{figure}

In this section we construct $CSB$ potential employing the one-boson exchange $(OBE)$ model
considering the in-medium meson propagators and the in-medium spinors for external nucleon legs 
as shown in the relevant Feynman diagrams in Fig.~\ref{csb}. To construct the $CSB$ potential one 
should first calculate the nucleon-nucleon scattering amplitude $\mathcal{M}^{(NN)}_{\pi\eta}(q^2)$ 
using the Fig.~\ref{csb}: 
\vspace*{-0.1cm}
\bea
\mathcal{M}^{(\!NN\!)}_{\pi\eta}(\!q^2\!)
\!\!\!&=&\!\!\!\left\{\!\bar{U}_3(P_3,s_3)\tau_3(1\!)\Gamma_{\pi}(q^2\!)U_1(P_1,s_1)\!\right\} \nn \\
\!\!\!&\times&\!\!\! \widetilde{D}_{\pi}(q^2)\Pi_{\pi\eta}(q^2)\widetilde{D}_{\eta}(q^2) \nn \\
\!\!\!&\times&\!\!\!\left\{\!\bar{U}_4(P_4,s_4)\Gamma_{\eta}(-q^2\!)U_2(P_2,s_2)\!\right\} \nn \\
\!\!\!&+&\!\!\!\left\{\!\bar{U}_3(P_3,s_3)\Gamma_{\eta}(q^2)U_1(P_1,s_1)\!\right\} \nn \\
\!\!\!&\times&\!\!\!\widetilde{D}_{\eta}(q^2)\Pi_{\pi\eta}(q^2)\widetilde{D}_{\pi}(q^2) \nn \\
\!\!\!&\times&\!\!\!\left\{\!\bar{U}_4(P_4,s_4)\tau_3(2\!)\Gamma_{\pi}(-q^2\!)U_2(P_2,s_2)\!\right\}. 
\label{nn:scat:amp}
\eea

In the above expression $U_l(P_l,s_l)$ where $l=1,\cdots,4$ represents in-medium spinors
for the external nucleon legs as represented by solid straight lines in Fig.~\ref{csb},
with four momentum $P_l$ and nucleon spin $s_l$. The meson-nucleon-nucleon interaction vertices 
are labeled by $1$ and $2$. The isospin operator $\tau_3$ takes care of the fact that only the neutral 
pion couples with the nucleon.
\vspace*{-0.5cm}

The potential in the momentum space is obtained from the scattering amplitude given in 
Eq.~(\ref{nn:scat:amp}) by substituting $q^2_0 =0$: 
\be
 V^{(NN)}_{CSB}(\vq^2) = \mathcal{M}^{(NN)}_{\pi\eta}(0,\vq^2)~, \label{mom:potential01}
\ee
along with the non-relativistic form of the spinors obtained by suitable expansion of the nucleon energy
$\en_N=\sqrt{\mn^2_N + {\bf p}^2_l}$ as
\bea
U({\bf p}_l,s_l) \simeq \left(1-\frac{{\bf P}^2}{8\mn^2_N}-\frac{{\bf q}^2}{32\mn^2_N}\right)
\left(\begin{array}{c}
1 \\
\frac{\sigma\cdot \left({\bf P} + {\bf q}/2\right)}{2\mn_{N}} \\
\end{array}
\right),
\label{spinor}
\eea

in terms of the average nucleon momentum defined by ${\bf P} =  ({\bf P}_1 + {\bf P}_3)/2=
({\bf P}_2 + {\bf P}_4)/2$ and, the three momentum of meson, ${\bf q} = {\bf p}_1-{\bf p}_3=
{\bf p}_2-{\bf p}_4$. Such expansion of the nucleon energy in the spinors helps one to simplify 
the spin structure of the $NN$ potential. ${\bf \sigma}$ represents the Pauli spin matrices.  
\vspace*{0.4cm}

Since the mixing amplitude contains both the vacuum and medium contributions, the $CSB$ potential
also contains two parts, namely a vacuum part $V^{(NN)}_{CSB,v}(\vq^2)$ and a medium part
$V^{(NN)}_{CSB,m}(\vq^2)$:
\be
V^{(NN)}_{CSB}(\vq^2)=V^{(NN)}_{CSB,v}(\vq^2)+V^{(NN)}_{CSB,m}(\vq^2).
\ee

It is important to note that mesons and nucleons are not point like particles. They have quark
structures. One must include the form factors at each nucleon-nucleon-meson interaction vertex while 
derive the potential within the framework of $OBE$ model \cite{Cohen95,Gardner96}. In the most of the 
calculations phenomenological form factors either monopole type \cite{Saito03,Abhee97,Cardarelli97}
or dipole type \cite{Machleidt01} are used at each vertex. Such form factors are obtained from the 
phenomenological fit of the two nucleon data. However, the vertex factor ought to be calculated 
from the same theory that provides the propagator \cite{Cohen95}. The form factor also cures the
problem of contact term in $OBE$ potential \cite{Cardarelli97}. 
  
In the present calculation we use phenomenological form factors of dipole type in space like region 
at each vertex. At each vertex the coupling constant ${\rm g}_a$ in Eq.~(\ref{mom:potential01}) is to 
be replaced as follows: 
\be
{\rm g}_a \rightarrow {\rm g}_a \left(\frac{\Lambda^2_a - m^2_a}{\Lambda^2_a+\vq^2}\right)~.
\ee

The cut-off parameter $\Lambda_a$ in the $OBE$ model separates the short ranged effects of the nuclear force 
while the long ranged parts are handeled by the meson propagators \cite{Cohen95,Gardner96}. 

Once we have the potential in momentum space we can easily obtain it in the coordinate space by Fourier 
transformation. Thus Eq.~(\ref{mom:potential01}) reduces to
\be
V^{(NN)}_{CSB}(r)=\int\frac{d^3\vq}{(2\pi)^3}V^{(NN)}_{CSB}(\vq^2)exp(-\mathrm{i\vq\cdot\vr}). 
\label{cor:potential01}
\ee
\subsection{Pseudoscalar interaction}
Substituting the in-medium meson propagators, phenomenological form factors, non-relativistic 
spinors, mixing amplitude and keeping the terms of the order $\mathcal{O}({\vq}^2/\mn^2_N)$ 
with some algebraic calculation one may write the $CSB$ potential in momentum space as 
\bla
&V^{(NN)}_{CSB,v}(\vq^2)=T^{+}_3C_{\pi\eta}A_{1\pi\eta}(\sigma_1\cdot\vq)(\sigma_1\cdot\vq)
\frac{{\rm g}_\pi{\rm g}_\eta}{4\mn^2_N}&&\nn \\
&\times \left[ \frac{1}{M^2_\eta-M^2_\pi} \left\{ \frac{M^2_\pi\delta_{\pi\eta} }{\vq^2 + M^2_\pi}  
- \frac{M^2_\eta\delta_{\eta\pi}} {\vq^2 + M^2_\eta}\right\} \right.&&\nn \\ 
&+ \frac{1}{\Lambda^2_\eta-\Lambda^2_\pi} \left. 
\left\{\frac{\Lambda^2_\pi\bar{\delta}_{\pi\eta} }{\vq^2 + \Lambda^2_\pi}  
-\frac{\Lambda^2_\eta\bar{\delta}_{\eta\pi}} {\vq^2 + \Lambda^2_\eta}\right\} \right],&& 
\label{ps:vac:mom:potential}
\ela
and
\bla
&V^{(NN)}_{CSB,m}(\vq^2)=T^{+}_3C_{\pi\eta}(\sigma_1\cdot\vq)(\sigma_1\cdot\vq)
\frac{{\rm g}_\pi{\rm g}_\eta}{4\mn^2_N}&&\nn \\
&\times \left[ \frac{B_{0\pi\eta}}{M^2_\eta-M^2_\pi} \left\{ \frac{\delta_{\pi\eta} }{\vq^2 + M^2_\pi}  
- \frac{\delta_{\eta\pi}} {\vq^2 + M^2_\eta}\right\} \right.&&\nn \\ 
&+\frac{B_{0\pi\eta}}{\Lambda^2_\eta-\Lambda^2_\pi}  
\left\{\frac{\bar{\delta}_{\pi\eta} }{\vq^2 + \Lambda^2_\pi}  
-\frac{\bar{\delta}_{\eta\pi}} {\vq^2 + \Lambda^2_\eta}\right\}&& \nn \\  
&+\frac{\left(\frac{B_{0\pi\eta}}{8\mn^2_N}+2B_{1\pi\eta} \right) }{M^2_\eta-M^2_\pi} 
\left\{ \frac{M^2_\pi\delta_{\pi\eta}}{\vq^2 + M^2_\pi}  
- \frac{M^2_\eta\delta_{\eta\pi}} {\vq^2 + M^2_\eta}\right\} &&\nn \\
&+\frac{\left(\frac{B_{0\pi\eta}}{8\mn^2_N}+2B_{1\pi\eta}  \right) }{\Lambda^2_\eta-\Lambda^2_\pi}\left. 
\left\{ \frac{\Lambda^2_\pi\bar{\delta}_{\pi\eta}}{\vq^2 + M^2_\pi}  
- \frac{\Lambda^2_\eta\bar{\delta}_{\eta\pi}} {\vq^2 + M^2_\eta}\right\}\right].&&
\label{ps:med:mom:potential}
\ela
In the above equations $T^{+}_{3}=\tau_{3}(1)+\tau_{3}(2)$,
\vspace*{-0.2cm}
\be
C_{\pi\eta}=\frac{1}{(1-A_{1\pi}-2B_{1\pi})(1-A_{1\eta}-2B_{1\eta})}~,\label{ps:cpieta}
\ee
and
\vspace*{-0.2cm}
\bse
\label{delta:ab}
\bea
\delta_{ab}&=&\left(\frac{\Lambda^2_a-m^2_a}{\Lambda^2_a-M^2_a}\right)
\left(\frac{\Lambda^2_b-m^2_b}{\Lambda^2_b-M^2_a}\right)~,\\
\bar{\delta}_{ab}&=&\left(\frac{\Lambda^2_a-m^2_a}{\Lambda^2_a-M^2_a}\right)
\left(\frac{\Lambda^2_b-m^2_b}{\Lambda^2_a-M^2_b}\right)~.
\eea
\ese

One may easily obtain the coordinate space potential by Fourier transform of the momentum space 
potential once constructed. Thus from Eq.~(\ref{ps:vac:mom:potential}) and Eq.~(\ref{ps:med:mom:potential}), 
the vacuum and medium parts of the potentials in coordinate space, respectively are as 
follows: 
\bea
V^{(NN)}_{CSB,v}(r)\!\!\!&=&\!\!\! -T^{+}_{3}C_{\pi\eta}A_{1\pi\eta}\frac{{\rm g}_\pi{\rm g}_\eta }{48\pi\mn^2_N}  \nn \\
&\times&\!\!\!\left[\frac{M^5_\pi~\delta_{\pi\eta}~U(x_\pi)-M^5_\eta~\delta_{\eta\pi}~U(x_\eta)}{M^2_\eta-M^2_\pi}\right. \nn \\    
&+&\left.\!\!\!\frac{\Lambda^5_\pi~\bar{\delta}_{\pi\eta}~U(X_\pi)-\Lambda^5_\eta~\bar{\delta}_{\eta\pi}~U(X_\eta)}
{\Lambda^2_\eta-\Lambda^2_\pi} \right],~ \label{ps:vac:coord:potential}
\eea
and
\vspace*{-0.125cm}
\bla
&V^{(NN)}_{CSB,m}(r)= -T^{+}_{3}C_{\pi\eta}\frac{{\rm g}_\pi{\rm g}_\eta }{48\pi\mn^2_N} 
&& \nn \\
\times&\left[B_{0\pi\eta}\left\{\frac{M^3_\pi~\delta_{\pi\eta}~U(x_\pi)-M^3_\eta~\delta_{\eta\pi}~U(x_\eta)}
{M^2_\eta-M^2_\pi}\right.\right. &&\nn \\      
&+ \left.  \frac{\Lambda^3_\pi~\bar{\delta}_{\pi\eta}~U(X_\pi)-\Lambda^3_\eta~\bar{\delta}_{\eta\pi}~U(X_\eta)}
{\Lambda^2_\eta-\Lambda^2_\pi} \right\}&& \nn \\
&+ \left(\frac{B_{0\pi\eta}}{8\mn^2_N}+2B_{1\pi\eta}\right) 
\left\{\frac{M^5_\pi~\delta_{\pi\eta}~U(x_\pi)-M^5_\eta~\delta_{\eta\pi}~U(x_\eta)}
{M^2_\eta-M^2_\pi}\right. &&\nn \\      
&+\left. \left.  \frac{\Lambda^5_\pi~\bar{\delta}_{\pi\eta}~U(X_\pi)-\Lambda^5_\eta~\bar{\delta}_{\eta\pi}~U(X_\eta)}
{\Lambda^2_\eta-\Lambda^2_\pi} \right\}\right]~.&&\label{ps:med:coord:potential}
\ela
\vspace*{-0.5cm}
Note that $x_a=M_ar$ and $X_a=\Lambda_ar$. We denote \\
\bea
U(x)&=&\left[\!(\sigma_1\cdot\sigma_2)\!+\!S_{12}(\hat{\bf r})\!\!
\left(\!\!1\!+\!\frac{3}{x}\!+\!\frac{3}{x^2}\!\!\right)\!\right]\!\frac{e}{x}^{-x}~,\\
S_{12}(\hat{\bf r})&=&3(\sigma_1\cdot\hat{\bf r})(\sigma_2\cdot\hat{\bf r})-(\sigma_1\cdot\sigma_2)~.
\eea
\vspace*{-0.4cm}
\subsection{Pseudovector interaction}
Similar to the pseudoscalar interaction, the $CSB$ potential in momentum space for 
vacuum and medium part in pseudovector interaction can be written as 
\bla
&V^{(NN)}_{CSB,v}(\vq^2)=T^{+}_3C^{\prime}_{\pi\eta}A^{\prime}_{1\pi\eta}(\sigma_1\cdot\vq)(\sigma_1\cdot\vq)
\frac{{\rm g}_\pi{\rm g}_\eta}{4\mn^2_N}&&\nn \\
&\times \left[ \frac{1}{M^{\prime2}_\eta-M^{\prime2}_\pi} \left\{ \frac{M^{\prime2}_\pi\delta^\prime_{\pi\eta} }
{\vq^2 + M^{\prime2}_\pi}  
- \frac{M^{\prime2}_\eta\delta^{\prime}_{\eta\pi}} {\vq^2 + M^{\prime2}_\eta}\right\} \right.&&\nn \\ 
&+ \frac{1}{\Lambda^2_\eta-\Lambda^2_\pi} \left. 
\left\{\frac{\Lambda^2_\pi\bar{\delta}^{\prime}_{\pi\eta} }{\vq^2 + \Lambda^2_\pi}  
-\frac{\Lambda^2_\eta\bar{\delta}^{\prime}_{\eta\pi}} {\vq^2 + \Lambda^2_\eta}\right\} \right],&& 
\label{pv:vac:mom:potential}
\ela
and
\bla
&V^{(NN)}_{CSB,m}(\vq^2)=T^{+}_3C^{\prime}_{\pi\eta}(2B^{\prime}_{1\pi\eta})(\sigma_1\cdot\vq)(\sigma_1\cdot\vq)
\frac{{\rm g}_\pi{\rm g}_\eta}{4\mn^2_N}&&\nn \\
&\times \left[ \frac{1}{M^{\prime2}_\eta-M^{\prime2}_\pi} 
\left\{ \frac{M^2_\pi\delta^{\prime}_{\pi\eta} }{\vq^2 + M^{\prime2}_\pi}  
- \frac{M^{\prime2}_\eta\delta^{\prime}_{\eta\pi}} {\vq^2 + M^{\prime2}_\eta}\right\} \right.&&\nn \\ 
&+ \frac{1}{\Lambda^2_\eta-\Lambda^2_\pi} \left. 
\left\{\frac{\Lambda^2_\pi\bar{\delta}^{\prime}_{\pi\eta} }{\vq^2 + \Lambda^2_\pi}  
-\frac{\Lambda^2_\eta\bar{\delta}^{\prime}_{\eta\pi}} {\vq^2 + \Lambda^2_\eta}\right\} \right].&&
\label{pv:med:mom:potential}
\ela
where, 
\be
C^{\prime}_{\pi\eta}=\frac{1}{(1-A^{\prime}_{1\pi}-2B^{\prime}_{1\pi})
(1-A^{\prime}_{1\eta}-2B^{\prime}_{1\eta})}~,\label{ps:cpieta}
\ee
Note that $\delta^{\prime}_{ab}$ and $\bar{\delta}^{\prime}_{ab}$ can be obtained from Eq.~(\ref{delta:ab})
by replacing $M_{a}\rightarrow~M^{\prime}_{a}$ and $M_{b}\rightarrow~M^{\prime}_{b}$ . The Fourier 
transformation of Eq.~(\ref{pv:vac:mom:potential}) and Eq.~(\ref{pv:med:mom:potential})
give us the potentials in coordinate space.
\bea
V^{(NN)}_{CSB,v}(r)&=&-T^{+}_{3}C^{\prime}_{\pi\eta}A^{\prime}_{1\pi\eta}\frac{{\rm g}_\pi{\rm g}_\eta }{48\pi\mn^2_N} 
\times \nn \\
&&\left[\frac{M^{\prime5}_\pi~\delta^{\prime}_{\pi\eta}~U(x^{\prime}_\pi)-
M^{\prime5}_\eta~\delta^{\prime}_{\eta\pi}~U(x^{\prime}_\eta)}
{M^{\prime2}_\eta-M^{\prime2}_\pi}\right. \nn \\    
&+&\left.  \frac{\Lambda^5_\pi~\bar{\delta}^{\prime}_{\pi\eta}~U(X_\pi)-\Lambda^5_\eta~\bar{\delta}^{\prime}_{\eta\pi}~U(X_\eta)}
{\Lambda^2_\eta-\Lambda^2_\pi} \right],~ \label{pv:vac:coord:potential}
\eea
and
\bea
V^{(NN)}_{CSB,m}(r)&=&-T^{+}_{3}C^{\prime}_{\pi\eta}(2B^{\prime}_{1\pi\eta})\frac{{\rm g}_\pi{\rm g}_\eta }{48\pi\mn^2_N} 
\times \nn \\
&&\left[\frac{M^{\prime5}_\pi~\delta^{\prime}_{\pi\eta}~U(x^{\prime}_\pi)-
M^{\prime5}_\eta~\delta^{\prime}_{\eta\pi}~U(x^{\prime}_\eta)}
{M^{\prime2}_\eta-M^{\prime2}_\pi}\right. \nn \\    
&+&\left.  \frac{\Lambda^5_\pi~\bar{\delta}^{\prime}_{\pi\eta}~U(X_\pi)-\Lambda^5_\eta~\bar{\delta}^{\prime}_{\eta\pi}~U(X_\eta)}
{\Lambda^2_\eta-\Lambda^2_\pi} \right],~ \label{pv:med:coord:potential}
\eea
where, $x^\prime_a=M^\prime_ar$.

\section{Result and discussion}
\label{result}
In this section we discuss our numerical results. To generate numerical 
data following meson parameters \cite{Machleidt87} listed in table \label{mesonpara} 
have been used. The nucleon mass in nuclear medium is estimated by solving Eq.~(\ref{nm})
self-consistently. We borrowed the values of coupling constant $(\mathrm{g}_\sigma)$
and mass $(m_\sigma)$ of the scalar meson $\sigma$ from Ref.~\cite{Serot86} to calculate 
the effective nucleon mass $(\widetilde{M}_N)$ in nuclear matter. 

\begin{table}[htb!]
\caption{Meson parameters.}
\begin{ruledtabular}
\label{mesonpara}
\begin{tabular}{lccc} 
         Mesons &  $m_a$ (MeV) & $\frac{\g^2_a}{4\pi}$ & $\Lambda_a$ (MeV) \\ \hline
  $\pi$     & $138.6$     & $14.6$  & $1300$    \\
  $\eta$   & $548.0$        & $5.0$    & $1500$    \\
\end{tabular}
\end{ruledtabular}
\end{table}

All the Figures in this section represent the difference of $CSB$ potentials 
between $nn$ and $pp$ systems in the coordinate space in $^1S_0$ state. We denote 
$\Delta{V} = V^{(nn)}_{CSB} - V^{(pp)}_{CSB}$ where, 
$V^{(NN)}_{CSB} = V^{(NN)}_{CSB,v} + V^{(NN)}_{CSB,m}$. We consider the nuclear matter 
density $\rho_0~(0.148~\mathrm{fm}^{-3})$. 
 
\begin{figure}[htb!]
\begin{center}
\includegraphics[scale=0.5,angle=0]{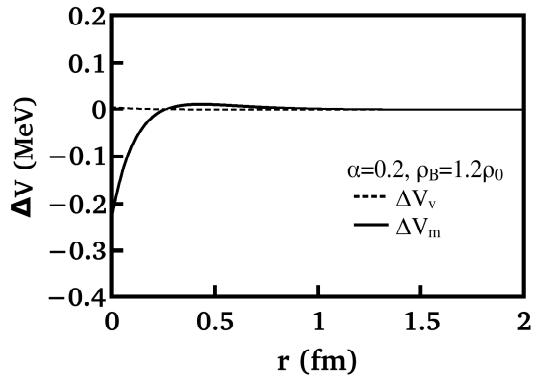}
\caption{The variation of $\Delta{V}_v$ and $\Delta{V}_m$ with $r$ for $PS$ 
interaction considering the in-medium $\pi$ and $\eta$ meson propagators.}
\label{fig:psvacmed}
\end{center}
\end{figure}

\begin{figure}[htb!]
\begin{center}
\includegraphics[scale=0.5,angle=0]{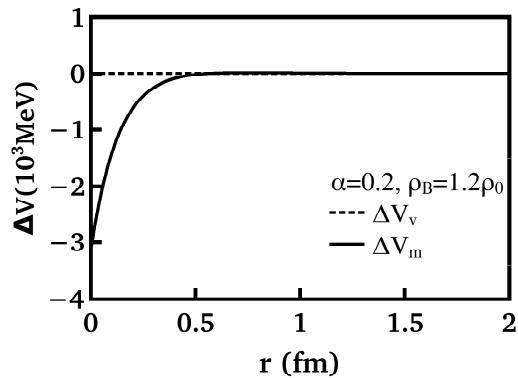}
\caption{The variation of $\Delta{V}_v$ and $\Delta{V}_m$ with $r$ for $PV$ 
interaction considering the in-medium $\pi$ and $\eta$ meson propagators.}
\label{fig:pvvacmed}
\end{center}
\end{figure}

At first we present the vacuum contribution, $\Delta{V}_v$ and medium contribution, 
$\Delta{V}_m$ for $PS$ interaction in Fig.~\ref{fig:psvacmed} and the same for $PV$
interaction in Fig.~\ref{fig:pvvacmed}, respectively. These figures show the variation 
of vacuum and medium parts with distance $r$ at density $\rho_B=1.2\rho_0$ and asymmetry  
$\alpha = 0.2$. The vacuum contribution is found to dominate over the medium 
contribution below $r \sim 0.25~\mathrm{fm}$ for $PS$ interaction while for $PV$ 
interaction the same effect is observed below $r \sim 0.5~\mathrm{fm}$. 
In addition, the medium contribution of $PS$ interaction is found $\sim 10^3$
times larger than that of $PV$ interaction below $r \sim 0.5~\mathrm{fm}$, however 
the vacuum contributions for both cases are comparable. 

\begin{figure}[htb!]
\begin{center}
\includegraphics[scale=0.5,angle=0]{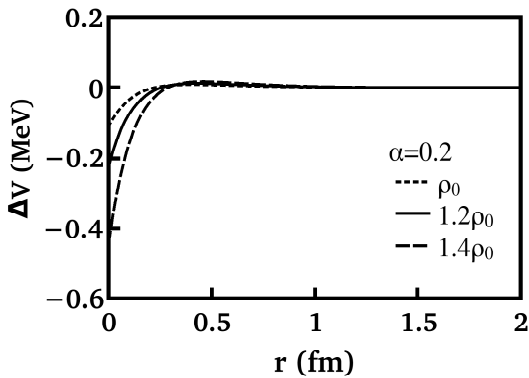}
\caption{The plot shows variation of $\Delta{V}$ with $r$ at different densities 
for $PS$ interaction considering the in-medium $\pi$ and $\eta$ meson propagators.}
\label{fig:psdens}
\end{center}
\end{figure}

\begin{figure}[htb!]
\begin{center}
\includegraphics[scale=0.5,angle=0]{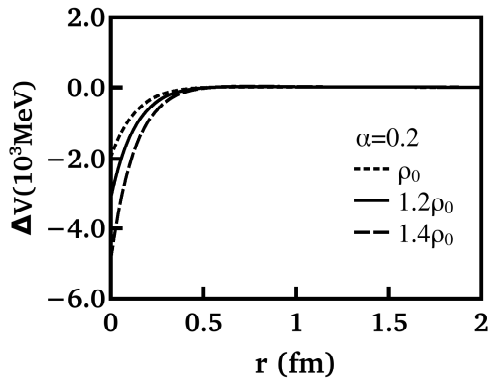}
\caption{The plot shows variation of $\Delta{V}$ with $r$ at different densities 
for $PV$ interaction considering the in-medium $\pi$ and $\eta$ meson propagators.}
\label{fig:pvdens}
\end{center}
\end{figure}

Figures \ref{fig:psdens} and \ref{fig:pvdens} show the difference of $CSB$ potentials
{\em i.e.} $\Delta{V}$ for $PS$ and $PV$ interactions at different densities but the 
same asymmetry $\alpha = 0.2$. The dotted, solid and dashed curves represent
$\Delta{V}$ at densities $\rho_B = \rho_0$, $1.2\rho_0$ and $1.4\rho_0$
respectively. 

It is observed that below $r \sim 0.5$ fm, $\Delta{V}$ for $PS$ interaction is  
$\sim 10^3$ times larger than that of $PV$ interaction as shown in Fig.~\ref{fig:pvdens} 
and Fig.~\ref{fig:psdens}. The origin of such large value possibly arises from the 
in-medium meson propagators. For $PS$ interaction $C_{\pi\eta}$ is found $\sim 10^{-5}$
times of $C^\prime_{\pi\eta}$ for $PV$ interaction. And this causes large contribution
of $CSB$ for $PS$ interaction compared to $PV$ interaction. It is also evident from the 
Fig.~\ref{fig:pvdens} and Fig.~\ref{fig:psdens} that higher the density higher is the 
$\Delta{V}$ below $r \sim 0.25 \mathrm{fm}~(0.5 \mathrm{fm})$ for $PS~(PV)$ interaction.     

\begin{figure}[htb!]
\begin{center}
\includegraphics[scale=0.5,angle=0]{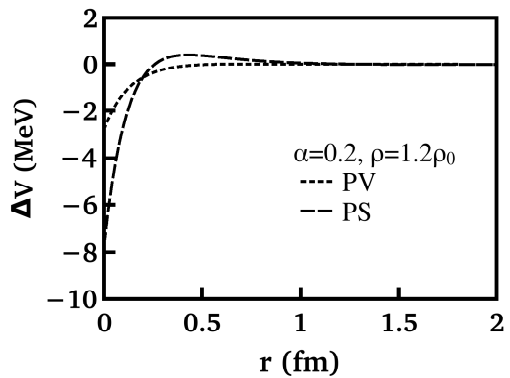}
\caption{The variation of $\Delta{V}$ with $r$ at $\alpha=0.2$ and $\rho_B=1.2\rho_0$
for $PS$ (dashed curve) and $PV$ (dotted curve) interactions is shown. The effective $\pi$ 
and $\eta$ meson masses are used to generate this graph, instead of propagators.}
\label{fig:pspvefm}
\end{center}
\end{figure}

In Fig.~\ref{fig:pspvefm} we show the variation of $\Delta{V}$ with $r$ both for 
$PS$ and $PV$ interactions considering the effective masses of $\pi$ and $\eta$ mesons
instead of in-medium propagators. To study the role of effective mass of meson we simply 
replaced the bare masses with their effective masses to the meson propagators. It is 
clear from Fig.~\ref{fig:pspvefm} that such replacement of bare mass with in-medium 
mass of meson makes $\Delta{V}$ comparable for both $PS$ and $PV$ interactions. 
At $r\sim 0.2$ fm, $\Delta{V}$s are found to be equal for both the cases and $\Delta{V}$
is positive between $r \sim 0.25 - 0.75$ fm for $PS$ interaction.      

\section{Summary}
\label{summary}
In the present work we have revisited $CSB$ due to $\pi$-$\eta$ mixing in nuclear 
matter employing the $OBE$ model and constructed the two-body $CSB$ potential which 
is class $III$ type. The in-medium nucleon propagator is used to calculate the meson 
self-energies and $\pi$-$\eta$ mixing amplitude and both calculations are restricted 
to the one loop order. We used these meson self-energies to obtain the in-medium 
meson propagators by solving the Swinger-Dyson equation. The bare propagators in 
the $CSB$ potential in momentum space is replaced by the in-medium $\pi$ and $\eta$
propagators.  

Furthermore, we also used the mixing amplitude in space like region to construct 
the two-body $CSB$ potential instead of constant or on-shell mixing amplitude 
and the effect of external nucleon legs is also taken into account. 

We noticed that the difference of $CSB$ $nn$ and $pp$ potentials, $\Delta{V}$
for $PV$ interaction is insignificant compared to that for $PS$ interaction while 
the effective masses of $\pi$ and $\eta$ instead of their in-medium propagators
shows comparable contributions for both the cases. 

\section*{ACKNOWLEDGMENT}
The author thanks Prof. Pradip K. Roy, Saha Institute of Nuclear Physics, and Mahatsab 
Mandal, Government General Degree College of Kalna, East Bardwan, India for their 
valuable comments and suggestions.

\appendix
\section{}
\label{appndixa}
Integrating over the azimuthal angle $\phi$ Eq.~(\ref{ps:med:selfen02}) reads
\bea
\Pi^{(N)}_{ab,m}(q^2)&\approx&2\left({\g_a}\over{2\pi}\right)\left({\g_b}\over{2\pi}\right)
{\int^{k_N}_0}\frac{\vk^2d\vk}{\sqrt{\mn^2_N+\vk^2}}\nn \\
&\times&{\int^{+1}_{-1}}dx\left[1+{q^4\over{4(k.q)^2}} \right] \nn\\
&=&2\left({\g_a}\over{2\pi}\right)\left({\g_a}\over{2\pi}\right) \left(I_1+I_2 \right),  \label{ps:med01}
\eea
where $x=\cos\theta$,
\bea
I_1\!\!&=&\!\!2\int^{k_N}_0\!\!\!\frac{\vk^2dk}{\sqrt{\mn^2_N+\vk^2}} \nn \\
&=&k_N\en_N-{1\over2}\mn^2_N\ln\left(\frac{\en_N+k_N}{\en_N-k_N}\right)~, \label{ps:med02}
\eea
and
\bea
I_2\!\!\!&=&\!\!\!{q^4\over4}\!\!\!\int^{k_N}_0\!\!\!\!\frac{\vk^2d\vk}{\sqrt{\mn^2_N+\vk^2}}
\!{\int^{+1}_{-1}}\!\!\!\frac{dx}{(q_0\en_N-|\vk||\vq|x)^2} \nn \\
\!\!&=&\!\!{q^4\over4}\!\!\!\int^{k_N}_0\!\!\!\frac{\vk^2d\vk}{\sqrt{\mn^2_N+\vk^2}}
\left[\frac{2}{(q_0\en_N)^2-(|\vk||\vq|)^2}\right] \nn \\
&=&{q^4\over{4q^2_0}}\int^{k_N}_0\frac{\vk^2d\vk}{\sqrt{\mn^2_N+\vk^2}}
\left[{2\over{\mn^2_N+z\vk^2}}\right] \nn \\
&\approx&\!\!\!{q^4\over{4q^2_0\mn^2_N}}\!\!\!\int^{k_N}_0\!\!\!\frac{\vk^2d\vk}{\sqrt{\mn^2_N+\vk^2}}
\left[2-{2z\over\mn^2_N}\vk^2\right]~,\label{ps:med03}
\eea
where, $z=1-\vq^2/q^2_0$. Note that $z<1$ and $z\vk^2<<\mn^2_N$. Thus we neglected
the term proportional to $z$ in Eq.~(\ref{ps:med03}).
\bea
I_2\!\!\!&\approx&\!\!\!{q^4\over{4q^2_0\mn^2_N}}I_1~.\label{ps:med04}
\eea
The terms proportional to $\vq^2/q^4_0$ have been neglected as $\vq^2/q^4_0<<1$ and 
$q^4/q^2_0$ is approximated to $(q^2_0-2\vq^2)$. Thus, Eq.~(\ref{ps:med04}) can be written as
\bea
I_2\!\!\!&\approx&\!\!\!(q^2_0\!-\!2\vq^2){1\over4}
\!\!\left[\! \frac{k_N\en_N}{\mn^2_N}\!-\!{1\over2}\!
\ln\!\left(\!\!\frac{\en_N\!+\!k_N}{\en_N\!-\!k_N}\!\!\right)\!\right]~.\label{ps:med06}
\eea
Substituting $I_1$ and $I_2$ in Eq.~(\ref{ps:med01}) we obtain the contribution of nucleon loop to
medium part Eq.~(\ref{ps:med:selfen03}). 
\section{}
\label{appendixb}
To remove the diverging part from Eq.~(\ref{pv:vac:selfen01}) we use simple subtraction method 
\cite{Biswas08}. Let us denote 
\bla
&\Pi^{(N)}_{v}(q^2)\!\!=\!\!\left(\!\frac{\mn_N}{M_N}\!\right)^2
\!\!\left[- 2 -\frac{1}{\epsilon}+\gamma_E+\ln(4\pi\mu^2)+2\ln(\mn_N)  \right. &&\nn\\ 
&+\left.2\frac{\sqrt{4\mn^2_N-q^2}}{q}\tan^{-1}\left(\frac{q}{\sqrt{4\mn^2_N-q^2}} \right) \right]~, &&
\label{appendixb:vac:01}
\ela
and
\be
\widetilde{\Pi}^{(N)}_{v}(q^2) = \Pi^{(N)}_{v}(q^2) - \Pi^{(N)}_{v}(q^2=m^2_a). 
\label{appendixb:vac:02}
\ee 
This will remove the divergence yielding the finite vacuum part of the self-energy:
\be
\Pi^{\prime(N)}_{aa,v}(q^2) = \left(\frac{\mathrm{g}_a}{2\pi}\right)^2 \widetilde{\Pi}^{(N)}_{v}(q^2)~q^2.
\ee

\end{document}